\documentclass[fleqn,twoside]{article}
\usepackage{espcrc2}

\usepackage{amsmath,graphicx}

\newcommand{\Lx}{\left(}
\newcommand{\Rx}{\right)}
\newcommand{\Li}[1]{{\mathop{\rm Li}_{#1}\nolimits}}
\newcommand{\opo}{\tilde {\cal O}}

\newcommand{\coeff}{\tilde C}
\title{Higgs Boson Production at Hadron Colliders}

\author{W.B. Kilgore\address{Physics Department, Building 510A, \\
        Brookhaven National Laboratory, Upton, New York 11973, USA}%
        \thanks{Supported by the U.~S.~Department of Energy
        under Contract No.~DE-AC02-98CH10886.}
 \\[-72pt]\hfill BNL-HET-02/18, hep-ph/0208143\\[61pt]
}
       
\begin{document}

\begin{abstract}
I report on a calculation of the inclusive Higgs boson production
cross section at hadron colliders at next-to-next-to-leading order in
QCD.  The result is computed as an expansion about the threshold
region.  By continuing the expansion to very high order, we map the
result onto basis functions and obtain the result in closed analytic
form.
\vspace{1pc}
\end{abstract}

\maketitle
\section{Introduction}
At the LHC, gluon fusion will be the most important mechanism for
Higgs boson production and discovery for masses below $\sim700$ GeV.
The discovery of the Higgs boson and the subsequent study of its
properties therefore relies on a solid theoretical understanding of
the gluon fusion production mechanism.  Unfortunately, next-to-leading
order (NLO) studies~\cite{nlo} of inclusive Higgs production do not
provide this solid understanding.  The NLO corrections are so large
(of order $70-100$\%) that one cannot assume that they provide a
reliable estimate of the total cross section.  The unsettled nature of
such an important signal clearly calls for a renewed effort to bring
this process under control.

Earlier this year, we computed the full NNLO corrections to the
hadronic cross section for Higgs boson
production~\cite{Harlander:2002wh} as an expansion about the threshold
region.  The expansion was carried out to very high ($18$th!) order so
that any uncertainty due to uncalculated higher terms would be very
small.  Very recently, this calculation has been confirmed by an exact
NNLO calculation of the partonic cross
section~\cite{Anastasiou:2002yz}.  Shortly before this conference
began, we extended our calculation of the expansion to sufficiently
high order as to allow us to invert the series and obtain the exact
result for the hadronic cross section.  In addition, we have used the
same methods to compute the NNLO corrections to pseudo-scalar Higgs
production, for which we present preliminary
results~\cite{Harlander:2002vv}.

\section{The Calculation}
In the limit that all quark masses except that of the top quark
vanish, gluons couple to Higgs only via top quark loops.  This
coupling can be approximated by an effective Lagrangian~\cite{efflag}\
corresponding to the limit $m_t\to \infty$, which is valid for a large
range of $M_H$, including the currently favored region between 100 and
200\ GeV.  The effective Lagrangian is
\begin{equation}
\label{eq::efflag}
{\cal L}_{\rm eff} = -\frac{H}{4v}C_1(\alpha_s)\,G_{\mu\nu}^aG^{a\,\mu\nu}\,,
\end{equation}
where $G_{\mu\nu}^a$ is the gluon field strength tensor, $H$ is the
Higgs field, $v\approx 246$\,GeV is the vacuum expectation value of
the Higgs field and $C_1(\alpha_s)$ is the Wilson coefficient, which
for this calculation we need to order $(\alpha_s^3)$~\cite{wilco}.

For the pseudo-scalar Higgs, we again assume that gluon fusion through
top-quark loops is the dominant production mechanism.  This assumption
is not valid in all models, especially when the ratio of vacuum
expectation values is large.  The effective Lagrangian for
pseudo-scalar production is~\cite{Chetyrkin:1998mw}
\begin{equation}
\begin{split}
  &{\cal L}_{Agg} = -g_t\,\frac{A}{v}\left[\coeff_1\, \opo_1
    + \coeff_{2}\,
    \opo_{2}\right]\,,\\
  &  \opo_1 = G^{a}_{\mu\nu}\tilde
  G^{a,\mu\nu}\,,\quad
  \opo_{2} = \partial_\mu\left(
  \sum_{q}\bar q \gamma^\mu\,\gamma_5 q\right)\,.
  \label{eq::leff}
\end{split}
\end{equation}
where $A$ is the pseudo-scalar Higgs field, $g_t$ is a model-dependent
coupling constant and $\tilde G^{a}_{\mu\nu}$ is the dual of the field
strength tensor:
\begin{equation}
\begin{split}
  \tilde G^{a}_{\mu\nu} = \epsilon_{\mu\nu\alpha\beta}\,G^{a,\alpha\beta}\,.
  \label{eq::gdual}
\end{split}
\end{equation}

The calculation breaks down into four contributions: virtual
corrections to two loops, single-real-emission to one loop,
double-real-emission at tree-level and mass factorization.  The
virtual, single-real and mass factorization terms are computed exactly
in closed analytic form.  The double-real contribution is by far the
hardest part of the calculation because of the complicated phase space
integrals.  We have computed this contribution by expanding the phase
space integration about the threshold limit, where the partonic
center-of-mass energy is close to the Higgs mass ($M_H^2/\hat{s}\equiv
x\to 1$).  Because Higgs production is dominated by threshold
corrections, the series expansion converges quite rapidly.  So, we
compute double-real-emission, and thus the partonic cross section, as
an expansion in $(1-x)$ and $\ln(1-x)$.
Note that if all coefficients are computed, this is an exact
expression.  In fact, one can obtain the exact result for the partonic
cross section from a finite number of terms.

\section{Inverting the series}
If one were to know the basis functions that make up the exact result
and one could expand the series out to enough terms, one could invert
the series, mapping it onto the basis functions.  One can obtain a
reasonable ansatz for the basis functions by examining the result for
Drell-Yan production~\cite{Hamberg:1991np}.
Using polylogagrithm identities~\cite{Lewin} this result can be
expressed in terms of functions which are analytic in
$(1-x)$ and powers of $\ln(1-x)$ times functions which are analytic in
$(1-x)$.  Each of these basis functions can be multiplied by a
pre-factor.  Again using the Drell-Yan result as a guide and knowledge
of the gluon splitting function, one can make an ansatz of the
possible pre-factors. The ans\"atze for pre-factors and basis
functions are shown in Table~\ref{table:funs}.

\begin{table}[htb]
\caption{Ansatz pre-factors and functions for exact result}
\label{table:funs}
\renewcommand{\tabcolsep}{.75pc} 
\renewcommand{\arraystretch}{2.0} 
\newcommand{\Dis}[1]{{$\displaystyle{#1}$}}
\begin{tabular}{c|ll}
\hline
\Dis{1}&
\Dis{1}				&\Dis{\ln(x)}\\
\Dis{\frac{1}{x}}&
\Dis{\ln^2(x)}			&\Dis{\ln^3(x)}\\
\Dis{\frac{1}{1-x}}&
\Dis{\Li2(1-x)}			&\Dis{\Li2(1-x)\ln(x)}\\
\Dis{\frac{1}{1+x}}&
\Dis{\Li2(1-x^2)}		&\Dis{\Li2(1-x^2)\ln(x)}\\
\Dis{1-x}&
\Dis{\Li3(1-x)}			&\Dis{\Li3\Lx-\frac{1-x}{x}\Rx}\\
\Dis{(1-x)^2}&
\Dis{\Li3(1-x^2)}		&\Dis{\Li3\Lx-\frac{1-x^2}{x^2}\Rx}\\
\Dis{(1-x)^3}&
\Dis{\Li3\Lx\frac{1-x}{1+x}\Rx}	&\Dis{\Li3\Lx-\frac{1-x}{1+x}\Rx}\\[5pt]
\hline
\end{tabular}
\end{table}

One sees that there are $7$ pre-factors and $14$ functions meaning
that if one can expand the series result out to $98$ terms, one can
map the result onto these $98$ functions.  The mapping can be verified
by computing still higher terms and comparing to the expansion of the
mapped functions.  To carry out this program, we have computed the
double-real radiation terms out to order $(1-x)^{100}$.  Since the
series starts at order $(1-x)^{-1}$, this gives us $102$ terms.

It turns out that the $1/x$ pre-factor never occurs in the cross
section (in Ref.~\cite{Hamberg:1991np} it appears in the Drell-Yan
correction term, not the Drell-Yan cross section) and the last two
basis functions, $\Li3((1-x)/(1+x))$ and $\Li3(-(1-x)/(1+x))$ always
occur together as the difference.
Thus, after the fact, we see that $78$ terms would have sufficed to
determine the functional form.  The additional $24$ terms
significantly over-determine the system and provide a strong
verification of the result.  In addition, we have compared our result
with that recently reported in Ref.~\cite{Anastasiou:2002yz}\ using
a completely independent method, and find exact agreement.

\section{Hadronic Results}
\begin{figure}[htb]
\includegraphics[width=\columnwidth]{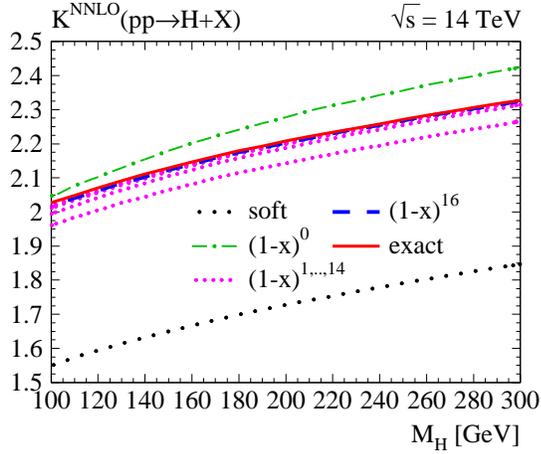}
\caption{NNLO cross section for Higgs production using the exact
partonic cross section and series expansions truncated at the
indicated values.}
\label{fig:convergence}
\end{figure}

\begin{figure}[htb]
\includegraphics[width=\columnwidth]{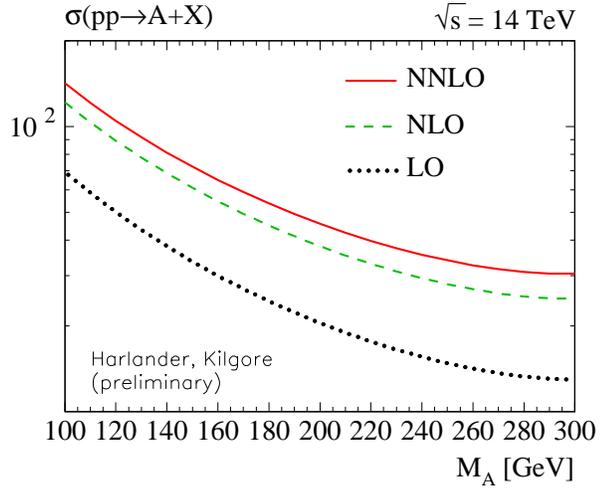}
\caption{Cross section for pseudo-scalar Higgs production at leading
order (LO), NLO and NNLO (preliminary)~\cite{Harlander:2002vv}.  The
model dependent couling $g_t$ has been set to unity.}
\label{fig:pseudo}
\end{figure}

In Ref.~\cite{Harlander:2002wh}, the partonic cross section was
computed using a series expansion out to order $(1-x)^{16}$.  The
difference between using that series expansion and the exact
calculation is quite small (less than $\sim1\%$) as one would expect.
In light of the intrinsic uncertainty due to scale dependence of order
$\pm10\%$, this difference is completely negligible.  The convergence
of the series for scalar Higgs production is shown in
Fig.~\ref{fig:convergence}.  Preliminary results for pseudo-scalar
Higgs production is shown in Fig.~\ref{fig:pseudo}

We obtain our hadronic results by folding the partonic cross section
with appropriate parton distribution functions (PDFs).  We use
approximate NNLO parton distribution functions~\cite{Martin:2002dr}
based on an approximation of the evolution
equation~\cite{structs,vanNeerven:2000wp}.


\begin{thebibliography}{99}
\expandafter\ifx\csname natexlab\endcsname\relax\def\natexlab#1{#1}\fi
\expandafter\ifx\csname bibnamefont\endcsname\relax
  \def\bibnamefont#1{#1}\fi
\expandafter\ifx\csname bibfnamefont\endcsname\relax
  \def\bibfnamefont#1{#1}\fi
\expandafter\ifx\csname citenamefont\endcsname\relax
  \def\citenamefont#1{#1}\fi
\expandafter\ifx\csname url\endcsname\relax
  \def\url#1{\texttt{#1}}\fi
\expandafter\ifx\csname urlprefix\endcsname\relax\def\urlprefix{URL }\fi
\providecommand{\bibinfo}[2]{#2}
\providecommand{\eprint}[2][]{\url{#2}}

\bibitem{nlo}
\bibinfo{author}{\bibfnamefont{S.}~\bibnamefont{Dawson}},
  \bibinfo{journal}{Nucl. Phys.} \textbf{\bibinfo{volume}{B359}},
  \bibinfo{pages}{283} (\bibinfo{year}{1991});\\
\bibinfo{author}{\bibfnamefont{A.}~\bibnamefont{Djouadi et~al.}},
  \bibinfo{journal}{Phys. Lett.} \textbf{\bibinfo{volume}{B264}},
  \bibinfo{pages}{440} (\bibinfo{year}{1991});\\
\bibinfo{author}{\bibfnamefont{D.}~\bibnamefont{Graudenz et~al.}},
  \bibinfo{journal}{Phys. Rev. Lett.} \textbf{\bibinfo{volume}{70}},
  \bibinfo{pages}{1372} (\bibinfo{year}{1993});\\
\bibinfo{author}{\bibfnamefont{D.}~\bibnamefont{Graudenz et~al.}},
  \bibinfo{journal}{Nucl. Phys.} \textbf{\bibinfo{volume}{B453}},
  \bibinfo{pages}{17} (\bibinfo{year}{1995}).

\bibitem{Harlander:2002wh}
\bibinfo{author}{\bibfnamefont{R.~V.} \bibnamefont{Harlander}}
  \bibnamefont{and} \bibinfo{author}{\bibfnamefont{W.~B.}
  \bibnamefont{Kilgore}}, \bibinfo{journal}{Phys. Rev. Lett.}
  \textbf{\bibinfo{volume}{88}}, \bibinfo{pages}{201801}
  (\bibinfo{year}{2002}).

\bibitem{Anastasiou:2002yz}
\bibinfo{author}{\bibfnamefont{C.}~\bibnamefont{Anastasiou et~al.}}
  (\bibinfo{year}{2002}), \eprint[http://arXiv.org/abs]{hep-ph/0207004}.

\bibitem{Harlander:2002vv}
\bibinfo{author}{\bibfnamefont{R.~V.} \bibnamefont{Harlander}}
  \bibnamefont{and} \bibinfo{author}{\bibfnamefont{W.~B.}
  \bibnamefont{Kilgore}} (\bibinfo{year}{2002}),
  \eprint[http://arXiv.org/abs]{hep-ph/0208096}.

\bibitem{efflag}
\bibinfo{author}{\bibfnamefont{J.} \bibnamefont{Ellis et~al.}},
  \bibinfo{journal}{Nucl. Phys.}
  \textbf{\bibinfo{volume}{B106}}, \bibinfo{pages}{292}
  (\bibinfo{year}{1976});\\
\bibinfo{author}{\bibfnamefont{A.} \bibnamefont{Vainshtein et~al.}},
  \bibinfo{journal}{Yad. Fiz.}
  \textbf{\bibinfo{volume}{30}}, \bibinfo{pages}{1368}
  (\bibinfo{year}{1979});
  \bibinfo{journal}{Usp. Fiz. Nauk}
  \textbf{\bibinfo{volume}{131}}, \bibinfo{pages}{537}
  (\bibinfo{year}{1980});\\
\bibinfo{author}{\bibfnamefont{M.} \bibnamefont{Voloshin}},
  \bibinfo{journal}{Yad. Fiz.} \textbf{\bibinfo{volume}{44}},
  \bibinfo{pages}{738} (\bibinfo{year}{1986}).

\bibitem{wilco}
\bibinfo{author}{\bibfnamefont{K.} \bibnamefont{Chetyrkin et~al.}},
  \bibinfo{journal}{Phys. Rev. Lett.} \textbf{\bibinfo{volume}{79}},
  \bibinfo{pages}{353} (\bibinfo{year}{1997});
  \bibinfo{journal}{Nucl. Phys.} \textbf{\bibinfo{volume}{B510}},
  \bibinfo{pages}{61} (\bibinfo{year}{1998});\\
\bibinfo{author}{\bibfnamefont{M.}~\bibnamefont{Kr\"amer et~al.}},
  \bibinfo{journal}{Nucl. Phys.} \textbf{\bibinfo{volume}{B511}},
  \bibinfo{pages}{523} (\bibinfo{year}{1998}).

\bibitem{Chetyrkin:1998mw}
\bibinfo{author}{\bibfnamefont{K.} \bibnamefont{Chetyrkin et~al.}},
  \bibinfo{journal}{Nucl. Phys.}
  \textbf{\bibinfo{volume}{B535}}, \bibinfo{pages}{3}
  (\bibinfo{year}{1998}).

\bibitem{Hamberg:1991np}
\bibinfo{author}{\bibfnamefont{R.}~\bibnamefont{Hamberg et~al.}},
  \bibinfo{journal}{Nucl. Phys.} \textbf{\bibinfo{volume}{B359}},
  \bibinfo{pages}{343} (\bibinfo{year}{1991}).

\bibitem{Lewin}
\bibinfo{author}{\bibfnamefont{L.}~\bibnamefont{Lewin}},
  \emph{\bibinfo{title}{Polylogarithms and Associated Functions}}
  (\bibinfo{publisher}{Elsevier North Holland, Inc.}, \bibinfo{address}{New
  York}, \bibinfo{year}{1981}).

\bibitem{Martin:2002dr}
\bibinfo{author}{\bibfnamefont{A.} \bibnamefont{Martin et~al.}},
  \bibinfo{journal}{Phys. Lett.}
  \textbf{\bibinfo{volume}{B531}}, \bibinfo{pages}{216}
  (\bibinfo{year}{2002}).

\bibitem{structs}
\bibinfo{author}{\bibfnamefont{S.} \bibnamefont{Larin et~al.}},
  \bibinfo{journal}{Nucl. Phys.}
  \textbf{\bibinfo{volume}{B427}}, \bibinfo{pages}{41}
  (\bibinfo{year}{1994});
  \bibinfo{journal}{Nucl. Phys.}
  \textbf{\bibinfo{volume}{B492}}, \bibinfo{pages}{338}
  (\bibinfo{year}{1997});\\
\bibinfo{author}{\bibfnamefont{A.}~\bibnamefont{Retey et~al.}},
  \bibinfo{journal}{Nucl. Phys.} \textbf{\bibinfo{volume}{B604}},
  \bibinfo{pages}{281} (\bibinfo{year}{2001}).

\bibitem{vanNeerven:2000wp}
\bibinfo{author}{\bibfnamefont{W.~L.} \bibnamefont{van Neerven et~al.}},
  \bibinfo{journal}{Phys. Lett.} \textbf{\bibinfo{volume}{B490}},
  \bibinfo{pages}{111} (\bibinfo{year}{2000}).

\end{thebibliography}

\end{document}